%% file: main.tex
\documentclass[conference]{IEEEtran}
\IEEEoverridecommandlockouts
\usepackage{etoolbox}
\usepackage{xstring}
\usepackage{multirow}
\DeclareListParser{\doslashlist}{/}
\newcounter{ndnNameComponentCounter}%
\newcommand{\name}[1]{{%
  \setcounter{ndnNameComponentCounter}{0}%
  \renewcommand{\do}[1]{{%
    \ifnumgreater{\value{ndnNameComponentCounter}}{0}{\allowbreak/}{}%
    \ifnumodd{\value{ndnNameComponentCounter}}{}{}%
    ##1}%
    \stepcounter{ndnNameComponentCounter}}%
``{\fontfamily{cmtt}\small\selectfont\IfBeginWith{#1}{/}{/}{}\doslashlist{#1}}''%
}}
\usepackage{color,soul}
\usepackage{xspace}

\makeatletter
\def\ps@IEEEtitlepagestyle{%
  \def\@oddfoot{\mycopyrightnotice}%
  \def\@evenfoot{}%
}
\def\mycopyrightnotice{%
  {\footnotesize  This paper has been accepted for publication by the 18th IEEE Annual Consumer Communications \& Networking Conference (CCNC). The copyright is with the IEEE. \hfill}% <--- Change here
  \gdef\mycopyrightnotice{}% just in case
}

\usepackage[hyphens]{url}
\usepackage[hidelinks]{hyperref}
\hypersetup{breaklinks=true}
\usepackage{cite}
\usepackage{amsmath,amssymb,amsfonts}

\DeclareMathOperator*{\argmin}{arg\,min}
\usepackage{breqn} %break lines automatically 
\usepackage{algorithmic}
\usepackage{graphicx}
\usepackage{subcaption}
\usepackage{textcomp}
\usepackage{xcolor}
\usepackage{enumitem}
\usepackage{caption}
\usepackage{amssymb}
\usepackage[font=small,labelfont=bf]{caption}
\def\BibTeX{{\rm B\kern-.05em{\sc i\kern-.025em b}\kern-.08em
    T\kern-.1667em\lower.7ex\hbox{E}\kern-.125emX}}
\begin{document}

\setlength{\columnsep}{0.215 in}
\def\BibTeX{{\rm B\kern-.05em{\sc i\kern-.025em b}\kern-.08em T\kern-.1667em\lower.7ex\hbox{E}\kern-.125emX}}

\newcommand{\spyros}[1]{\textcolor{black}{#1}}
\newcommand{\reza}[1]{\textcolor{black}{#1}}
\newcommand{\xin}[1]{\textcolor{black}{#1}}
\newcommand{\peggy}[1]{\textcolor{brown}{#1}}

\newcommand{\ie}{{\em i.e.,}\ }
\newcommand{\eg}{{\em e.g.,}\ }
\newcommand{\sol}{{\em DLWIoT}\xspace}

\title{\sol: Deep Learning-based Watermarking for Authorized IoT Onboarding}

\author{\IEEEauthorblockN{Spyridon Mastorakis}
\IEEEauthorblockA{\textit{Dept. of Computer Science} \\
\textit{University of Nebraska Omaha}\\
%City, Country \\
smastorakis@unomaha.edu}
\and
\IEEEauthorblockN{Xin Zhong}
\IEEEauthorblockA{\textit{Dept. of Computer Science} \\
\textit{University of Nebraska Omaha}\\
%City, Country \\
xzhong@unomaha.edu}
\and
\IEEEauthorblockN{Pei-Chi Huang}
\IEEEauthorblockA{\textit{Dept. of Computer Science} \\
\textit{University of Nebraska Omaha}\\
%City, Country \\
phuang@unomaha.edu}
\and
\IEEEauthorblockN{Reza Tourani}
\IEEEauthorblockA{\textit{Dept. of Computer Science} \\
\textit{Saint Louis University}\\
%%City, Country \\
reza.tourani@slu.edu}
}

%\author{
%    Spyridon Mastorakis,
%    Xin Zhong,
%    Pei-Chi Huang,
%    and Reza Tourani
	
%	\thanks{
%		S. Mastorakis, P. Huang, and X. Zhong are with the Department of Computer Science, University of Nebraska, Omaha, USA.
		
%		R. Tourani is with the Department of Computer Science, Saint Louis University, USA.
%	}	
%}

\maketitle

\begin{abstract}
\input{abstract}
\end{abstract}

\begin{IEEEkeywords}
Internet of Things (IoT), IoT onboarding, deep learning, watermarking
\end{IEEEkeywords}

\maketitle

\section{Introduction}
\input{introduction} 
\label{sec:intro}

%\section{Use-case: Image Watermarking Based on Deep Learning for IoT Onboarding}
%\label{sec:usecase}
%\input{usecase}

\section{\sol Threat Model and Design}
\label{sec:system}
\input{system}

% \section{Why Do We Need Compute Reuse?}
% \label{sec:motivation}
% \input{motivation}

\section{Deep Learning-based Image Watermarking}
\label{sec:deeplearning}
\input{deeplearning}

\section{Experimental Evaluation}
\label{sec:eval}
\input{evaluation}

%\section{Discussion}
%\label{sec:discussion}
%\input{discussion}

\section{Related Work}
\label{sec:related}
\input{related}

\section{Conclusion and Future Work}
\label{sec:conclusion}
\input{conclusion}

\section*{Acknowledgements}

This work is partially supported by a pilot award from the Center for Research in Human Movement Variability and the NIH (P20GM109090), the National Science Foundation under award CNS-2016714, a planning award from the Collaboration Initiative of the University of Nebraska system, and the Nebraska Tobacco Settlement Biomedical Research Development Funds.

\bibliographystyle{IEEEtran}
\bibliography{ref}

\end{document}

%% file: abstract.tex
The onboarding of IoT devices by authorized users constitutes both a challenge and a necessity in a world, where the number of IoT devices and the tampering attacks against them continuously increase. Commonly used onboarding techniques today include the use of QR codes, pin codes, or serial numbers. These techniques typically do not protect against unauthorized device access--a QR code is physically printed on the device, while a pin code may be included in the device packaging. As a result, any entity that has physical access to a device can onboard it onto their network and, potentially, tamper it (\eg install malware on the device). To address this problem, in this paper, we present a framework, called Deep Learning-based Watermarking for authorized IoT onboarding (\sol), featuring a robust and fully automated image watermarking scheme based on deep neural networks. \sol embeds user credentials into carrier images (\eg QR codes printed on IoT devices), thus enables IoT onboarding only by authorized users. Our experimental results demonstrate the feasibility of \sol, indicating that authorized users can onboard IoT devices with \sol within 2.5--3sec.

%Secure onboarding of IoT devices at scale remains a challenge. The most popular onboarding techniques are using quick response (QR) codes, pin code or serial number. However, these techniques have severely hindered their applicability, e.g., adding authentication mechanism to prevent users from data and information leakage. This paper proposes a novel and secure framework, called deep learning-based watermarking for authorized IoT onboarding (\sol), which enables only authorized users to onboard IoT devices onto their network through an image watermarking scheme based on deep neural networks.
%Our approach can simplify and accelerate time-consuming onboarding to IoT device management to eliminate one-off device staging or configuration. We evaluate the feasibility and \TODO{tradeoffs} of \sol from ene-to-end onboarding delay. The results indicate that \sol does not degrade the user quality of experience during the onboarding process, since it onboards IoT devices for authorized users in about 2.5 sec.

%% file: introduction.tex
%In a world of growing electronic and mobile devices equipped with screens and cameras, enabling screens and cameras to communicate has been attracting great attention. The information is encoded as an image which can be displayed on and surfaces where any camera equipped device can retrieve the information by capturing the image. One of the popular implementation is QR-Codes (Quick Response Codes)~\cite{ISOIEC18004}, where information, e.g., a URL or a short text, can be encoded as a 2D barcode.

%thomas2018multilevel

In a world, where the number of IoT devices rapidly increases every year, the onboarding of such devices has been a challenge, especially when security becomes a requirement. One of the common techniques for IoT onboarding today is the use of Quick Response (QR) codes~\cite{ISOIEC18004} (other popular out-of-band IoT onboarding techniques include the use pin codes or serial numbers~\cite{latvala2020evaluation}). A user scans the QR code of an IoT device with his/her mobile phone and this QR code is translated to a url, which allows the user to communicate with a server typically located on the cloud. Through this process, an IoT device acquires the necessary configuration and credentials, so that it onboards the user IoT network (\ie becomes an operational part of this network)~\cite{nour2020compute, mastorakis2020icedge}.

However, such onboarding techniques typically do not protect against unauthorized device use, since, for example, a QR code is printed on the device or a pin code may be inside the retail packaging of the device. As a result, unauthorized users or any entity that has physical access to the IoT device can onboard it onto their personal network and, potentially, tamper the device (\eg install malicious software on it).

To protect against unauthorized IoT onboarding, in this paper, we present Deep Learning-based Watermarking for authorized IoT onboarding (\sol). \sol is a trusted third-party service that interacts with the users and manufacturers for secure onboarding of IoT devices through an image watermarking scheme~\cite{cox2007digital} based on deep Neural Networks (NNs). 
%order their IoT devices from a trusted party (\eg manufacturer), while 
In our design, the users contact the \sol service after purchasing their IoT devices. \sol embeds a watermark instructed by the user (\eg user credentials) covertly onto a carrier image (\eg QR code). The marked image (the watermark embedded into the carrier image--\eg a QR code with the user credentials embedded to it) is printed on the device. 
%or is transmitted back to the user through a secure connection (\eg the image can be accessed by the user through a mobile application provided by the trusted party). 
When the user receives the ``watermarked'' IoT device, this image will be used for the on-boarding of the device onto the user network. Specifically, the user will take a picture of the image with his/her mobile phone%(or access the image through a mobile application)
, the watermark will be extracted, and the device will be onboarded. %Note that only the authorized user will be able to extract the watermark and onboard the device. 
Attackers will not be able to onboard and tamper the device even if they have physical access to it, assuming that they do not have access to the credentials used for the watermark creation. 

\subsection {Motivation and Contribution}

\begin{figure}[t]
\centering
 \begin{subfigure}[t]{0.43\columnwidth}
 \centering
 \includegraphics[width=\textwidth]{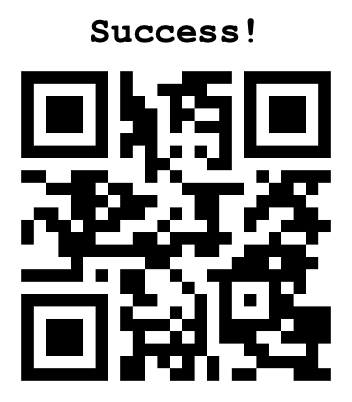}
 \caption{QR code without distortion}
 \label{qr}
 \end{subfigure}\hfil
 \begin{subfigure}[t]{0.43\columnwidth}
 \centering
 \includegraphics[width=\textwidth]{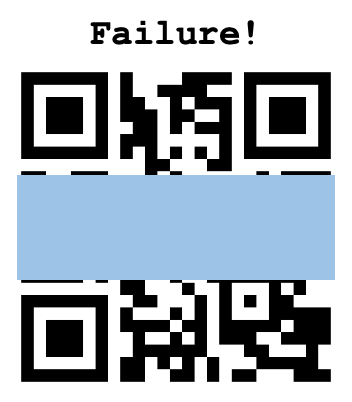}
 \caption{QR code with distortion of $\ge$ 33\%}
  \label{qr-33}
 \end{subfigure}\hfil
 %\vspace{-0.2cm}\caption{Test}% 100, 250, 500, and 1000}
 %\label{motivation}
 %\begin{subfigure}[b]{0.47\columnwidth}
 %\centering
 %\includegraphics[width=0.85\textwidth]{figs/qr-loc1.png}
 %\caption{Distorted QR code locator (top-left corner).}
 %\label{qr-loc1}
 %\end{subfigure}\hfil
 %\begin{subfigure}[b]{0.47\columnwidth}
 %\centering
 %\includegraphics[width=0.85\textwidth]{figs/qr-loc2.png}
 %\caption{Distorted QR code locator (bottom-left corner).}
 % \label{qr-loc2}
 %\end{subfigure}\hfil
 %\vspace{-0.2cm}
 \caption{Example of a successful and failed QR code scanning.}% 100, 250, 500, and 1000}
 \label{motivation}
 \vspace{-0.3cm}
\end{figure}

\noindent \textbf{Why using deep NNs for image watermarking in \sol:} Among the goals of an image watermarking system is robustness--the watermark must survive even after distortion or quality degradation of the marked image~\cite{cox2007digital}. This is especially needed when a marked image is printed on a device and/or the embedded watermark is extracted from images captured by a mobile phone. \spyros {In our use-case, in addition to attackers potentially tampering IoT devices, the attackers can further attempt to distort the marked images printed on the devices in order to prevent legitimate users from onboarding their devices. \sol can utilize any carrier image for IoT onboarding, but, in this paper, we primarily focus on QR codes given their simplicity and widespread use as the means to onboard IoT devices today. More specifically, as illustrated in Fig.~\ref{motivation}, legacy QR codes, even when error correction techniques are applied, can tolerate up to limited amounts of distortion--typically up to 15\%~\cite{qr-errors} and at most up to 33\%~\cite{stark2013qr}.}

% (Fig.~\ref{distortion})
%Moreover, if one of the QR code locators (rectangular boxes on the top right and left and bottom left corners) is distorted, the QR code scanning will fail.}

In \sol, a deep NN for watermarking can provide enhanced robustness during the process of embedding a watermark into a marked image and extracting the watermark from camera resamples of a marked image. \spyros {As a result, marked images (\eg QR codes) created through \sol can tolerate severe distortions (up to 85-90\%). More specifically,} the deep NN of \sol will dynamically learn the rules of watermark embedding and extraction, resulting in a fully automated and optimized watermarking system that is able to deal with different degrees of distortion and image quality degradation (\eg lighting variations, compression, cropping, lens distortion)~\cite{pramila2007camera}. On the other hand, conventional watermarking schemes are based on fixed watermark embedding and extraction algorithms, thus being able to tolerate only certain types of distortions. 

%\begin{figure}[t]
%\centering
% \begin{subfigure}[t]{0.42\columnwidth}
% \centering
% \includegraphics[width=0.85\textwidth]{figs/pic.png}
% \caption{Marked image without distortion}
% \label{qr}
% \end{subfigure}\hfil
% \begin{subfigure}[t]{0.42\columnwidth}
% \centering
% \includegraphics[width=0.85\textwidth]{figs/pic-dis.PNG}
% \caption{Marked image with distortion}
%  \label{qr-33}
% \end{subfigure}\hfil
% \caption{Example of a marked image with and without distortion.}
% \label{distortion}
% \vspace{-0.3cm}
%\end{figure}

\noindent \textbf{How is \sol different than setting a password for each IoT device:} \sol not only supports the embedding of any binary-encoded image into a marked image, \spyros {but it can also embed information, such as the user's voice (\eg a recorded word or phrase). In addition,} \sol can embed various biometrics, including users' fingerprints, iris scans, and gestures. Utilizing biometrics provides stronger security guarantees compared to password-based mechanisms~\cite{UluPanPra04} due to advanced social engineering and dictionary attacks against password-based systems as well as weak and predictable password selection. \spyros {To showcase the \sol design and capabilities, in this paper, we study the use-case of fingerprint scans to} represent user credentials due to its popularity and hardware availability on user devices. We evaluate the embedding of fingerprint scans into marked images in Section~\ref{sec:eval}.
%

%Finally, a deep NN can dynamically learn the rules of watermark embedding and extraction, resulting in a fully automated and optimized watermarking system. 
% during the process of embedding a watermark to a marked image and extracting the watermark from camera resamples of a marked image, being able to deal with different degrees of distortion and image quality degradation, such as lighting variations, compression, and lens distortion.

%Image watermarking is designed for authentication at its beginning~\cite{cox2007digital}. Among common factors including fidelity, capacity and robustness, image watermarking algorithms for covert communication often highlight robustness, \ie the watermark must survive even after distortion of the marked-image. This is especially needed when a marked-image is resampled by a phone camera, where the distortion can be a highly comprehensive combination of noise such as lighting variations, compression, quality degradation, and lens distortion~\cite{pramila2007camera}. To this end, a deep NN for blind watermarking has great potential to provide enhanced robustness without requiring prior knowledge of the distortion, and hence facilitating our target IoT application. 

\noindent \textbf {Contribution:} Our contribution is two-fold: (i) we present a novel scheme for image watermarking based on deep NNs that provides fully automated watermark embedding and extraction of enhanced robustness (Section~\ref{sec:deeplearning}); and (ii) we take advantage of this scheme for the design of \sol, a framework for authorized IoT onboarding, (Section~\ref{sec:system}) and we evaluate the feasibility and tradeoffs of \sol (Section~\ref{sec:eval}). Our experimental results indicate that \sol %does not degrade the user quality of experience during the onboarding process, since it 
is able to onboard IoT devices for authorized users in $2.5-3sec$.

%\noindent \textbf {Paper structure:} The remainder of the paper is structured as follows: Section~\ref{sec:system} presents our system design. Section~\ref{sec:deeplearning} describes \sol in detail. Section~\ref{sec:eval} describes the performance evaluation results and a real-life case study. Section~\ref{sec:discussion} describes \sol discussion. Section~\ref{sec:related} summarizes the related work. Section~\ref{sec:conclusion} concludes this work and discusses the future work.

%\xin{Reviewer Q1: explain the difference between the proposed methods with Xin's arXiv. Reviewer Q2: did not compare the wm method in this paper. we may not agree these should be in this paper (but added some sentences to answer these two questions). But the cause behind may be the reviewer wants to know more about the connection between \sol and watermark (he also asked "why using watermark"). We may solve these problems at the same time by: (1) stating the requirements of \sol. (2) the deep watermarking scheme satisfy these requirements.}

%% file: system.tex
\spyros {In this section, we first discuss the \sol threat model and assumptions, and we then present the design of the \sol architecture.}

\subsection {Threat Model and Assumptions}
\reza {As we mentioned earlier, common practices for onboarding an IoT device include scanning a QR code or using the device's serial number (or pin code) that is printed on the device. 
The potential threats from an unauthorized user with physical access to the device include: {\it (a)} onboarding the device into the unauthorized user's network aiming to tamper with the device (\eg installing malware such as Mirai~\cite{AnaCocDur17}); and {\it (b)} adding distortion or running a subtle cropping attack to prevent the legitimate user from successfully onboarding the device.
In this paper, we assume that the \sol service and IoT device manufacturers are trustworthy. This is a fair assumption since neither the \sol service nor manufacturers gain benefit by interfering with the onboarding process. }

\subsection {\sol Design}

\begin{figure}
    \centering
    \includegraphics[width=1\linewidth]{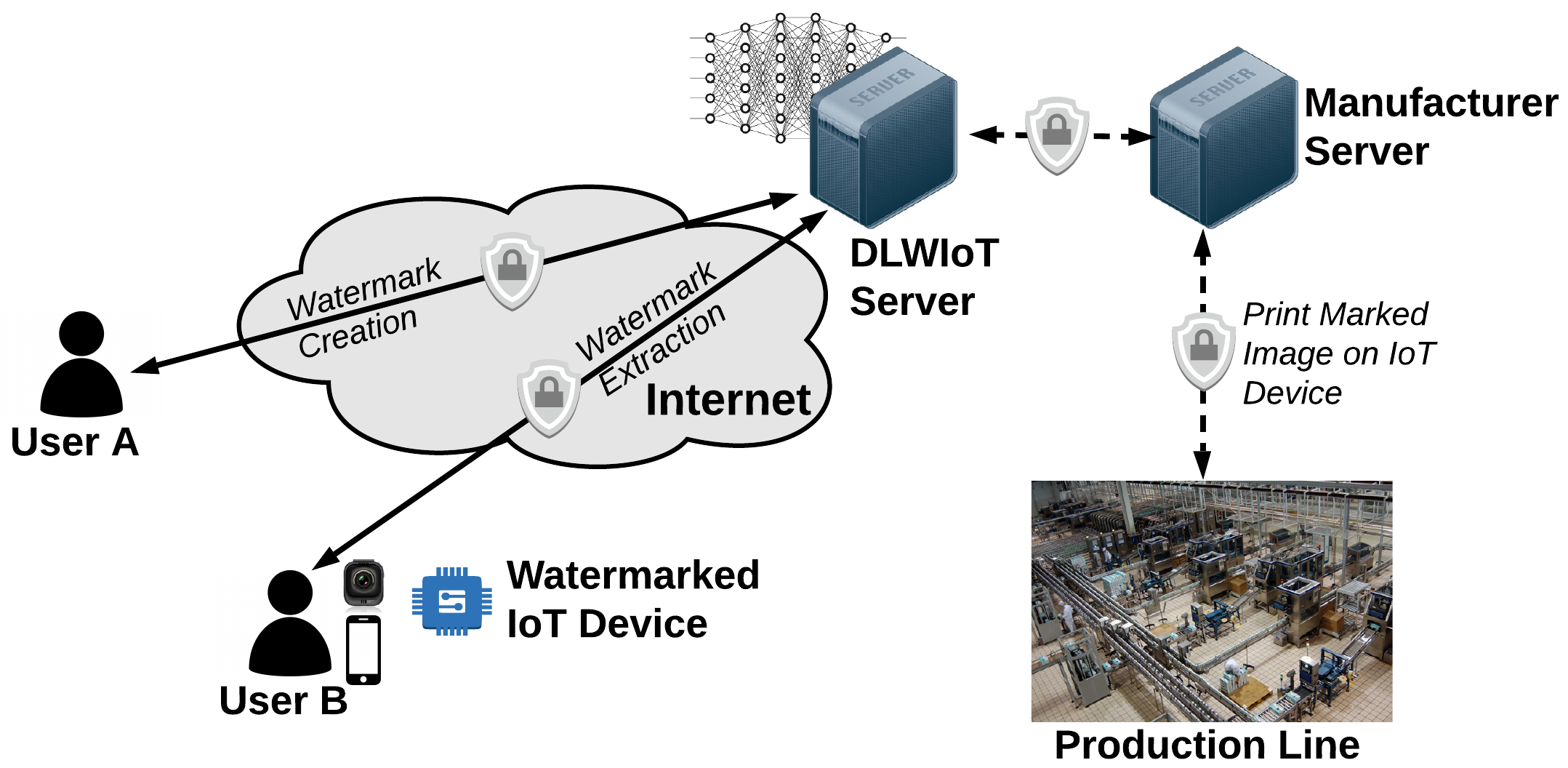}
    %\vspace{-0.5cm}
    \caption{\sol design and example workflow.}
    \vspace{-0.4cm}
    \label{fig:system}
\end{figure}

Figure~\ref{fig:system} illustrates the design of \sol service. In the context of \sol, we assume that a user buys an IoT device online (\eg through a manufacturer's website or an online shopping website). Once the user has completed her purchase, she will be redirected to the \sol, which is a trusted third-party service similar to a certificate authority. Through a secure connection (\eg SSL/TLS~\cite{dierks2008transport}), the user (\eg user A in our example) uploads her credential that she would like to use for the watermark creation and a carrier image to the \sol server. Alternatively, the user may select a carrier image (\eg a QR code) among the ones offered by the \sol server.
%

%The \sol service will be responsible for contacting a manufacturer's server to authenticate

%we assume that users purchase their IoT devices directly from trusted manufacturers. These manufacturers are willing to embed user-instructed watermarks (\eg user credentials) into carrier images (\eg QR codes) and print the resulting marked images on the produced IoT devices.

%After purchasing one or more IoT devices, users (\eg user A in our example) contact the \sol server. 

Once the user credentials are received by the \sol server, our embedding deep NN, running on the same or a different \sol server (illustrated as running on the same server for simplicity in Figure~\ref{fig:system}) will create a marked image by embedding the user credentials (\eg fingerprint scan, recorded voice, image, cryptographic key)\footnote{\spyros {Note that storing user biometrics on the \sol server may come with certain privacy concerns that should be considered. However, we would like to note that approaches to provide secure and privacy-preserving biometric storage and identification as a service on the cloud have been recently proposed~\cite{haghighat2015cloudid, barra2018cloud}. Such approaches can be utilized to complement the \sol design. To alleviate privacy concerns related to storing user biometrics on the \sol server, \sol allows users to utilize a wide variety of other credentials (\eg any image, any recorded word or phrase).}} into the carrier image. The \sol service securely transmits the marked image to the manufacturer server and, subsequently, the production line. Finally, the watermarked image will be printed on the IoT device before delivering it to the user.

Upon receiving the watermarked IoT device, the user (user B in Figure~\ref{fig:system}) takes a picture of the marked image on the device. Subsequently, the user securely sends the picture and the credentials embedded in the marked image (\eg fingerprint scan) to the \sol server (illustrated for simplicity as the same server for the watermark creation in Figure~\ref{fig:system}).
%
%{\color{blue} This picture is sent through a secure connection to a \sol server (illustrated for simplicity as the same server used for the watermark creation in Fig.~\ref{fig:system}) that runs our extracting deep NN along with a random pseudonym $p^i \in [p^{l}, p^{h}]$ used in the creation of the watermark.
%
%The server stores the users' pseudonym ranges in an ordered binary tree. 
%
%Then, given $p_i$ and the user's identify, the server makes ${\cal O}(\log |{\cal U}|)$ comparisons, where ${\cal U}$ is the set of users, to validate if the pseudonym belongs to that user for the onboarding process.
%}
%
The \sol server extracts the embedded credentials from the marked image and compares them against the user-provided credentials. Upon successful verification, the server sends the contact information (\eg IP address, TCP/UDP port, public key) of the onboarding server to the user.

%% file: deeplearning.tex
%Image watermarking refers to the process of embedding and extracting information covertly on a cover image. 
%\TODO{Check titles with Xin}
In this section, we present a deep learning-based image watermarking scheme, which is the core of \sol. 
\xin{To facilitate our IoT onboarding scenario, the deep NN is specially trained for an image watermarking with three main merits:} (i) to obtain an automated system without requiring domain knowledge, we exploit the fitting ability of deep NNs in learning image watermarking algorithms; (ii) we propose a deep learning architecture suitable to image watermarking that trains in an unsupervised manner to reduce human intervention; and (iii) the proposed scheme achieves robustness without any prior knowledge or adversarial examples of possible attacks.

%\subsection{Overall Design}

\subsection{Overall Architecture}
An image watermarking scheme often consists of watermark embedding and extracting stages, and each stage can be decomposed into several steps in typical methods. The watermark embedding stage aims to insert a watermark into a cover image. The first step is to project a cover image into one of its feature spaces in spatial, frequency, or other domains. The obtained feature space is then modified to carry the watermark. To create a marked image, the modified feature space is projected back into the cover image space. Inversely, watermark extraction is to project the marked image to the same feature space and then separate the watermark information. The watermark can be transformed or encoded based on different target applications. An image watermarking scheme often highlights its fidelity (\ie high similarity between the marked and the cover image) and robustness (\ie keeping the integrity of the watermark when there are noise and/or attacks applied to the marked image).

\begin{figure*}
    \centering
    \includegraphics[width=1\linewidth]{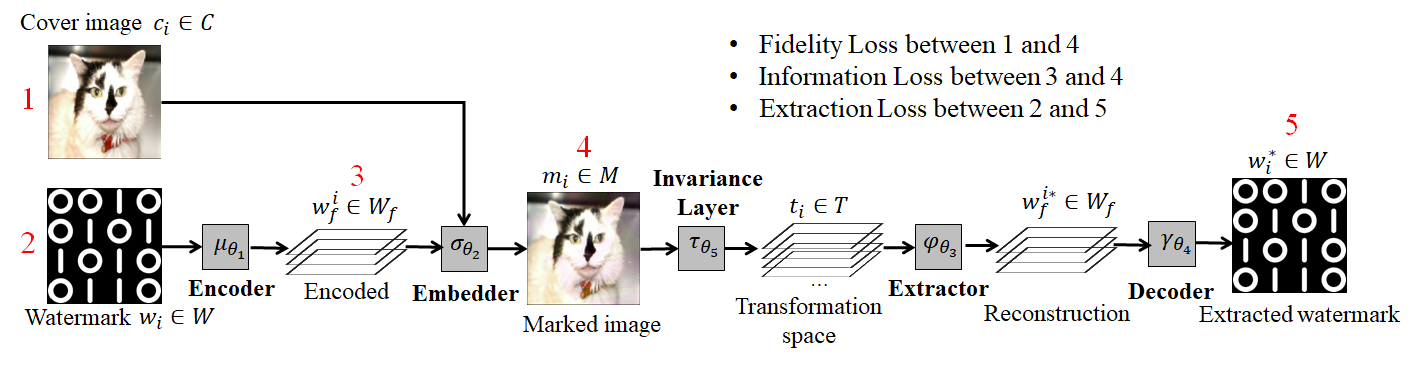}
    \vspace{-0.6cm}
    \caption{Overall architecture of the watermarking scheme.}
    \vspace{-0.4cm}
    \label{fig:overall_architecture}
\end{figure*}

The idea of the proposed scheme is to develop a deep learning model to learn and generalize image watermarking algorithms. As shown in Figure~\ref{fig:overall_architecture}, given two input spaces of watermark images and cover images, $W$ and $C$, we first fit a function that encodes $W$ to its encoded space $W_f$ with NN $\mu_{\theta_1}$ parameterized by $\theta_1$. Then, an embedder function that inserts $W_f$ into (a domain of) $C$ is fit by NN $\sigma_{\theta_2}$ parameterized by $\theta_2$. The obtained space after embedding for the marked image is named as $M$. To handle possible distortions, an NN $\tau_{\theta_5}$ parameterized by $\theta_5$ is introduced to fit a function coverting $M$ to its transformed space $T$. During the transformation, $\tau_{\theta_5}$ preserves information about $W_f$ while rejecting all irrelevant noise on $M$, and hence providing robustness to the proposed scheme. Finally, the inverse watermark reconstruction functions are fit by two NNs, $\varphi_{\theta_3}$ and $\gamma_{\theta_4}$ with trainable parameters $\theta_3$ and $\theta_4$, that extract $W_f$ from $T$ and decode $W$ from $W_f$ respectively. Note that the convolutional NNs applied in the proposed scheme not only fit the processes of feature extraction and feature space modification performed in traditional watermarking schemes, but also optimize these processes dynamically.

\subsection{Scheme Objective}
The entire architecture is trained as a single deep NN with several loss terms designed for image watermarking. Given the data samples $w_i \in W, i=1,2,3,... $ and $c_i \in C, i=1,2,3,... $, the proposed scheme can be trained in an unsupervised manner. There are two inputs $w_i$ and $c_i$, and two outputs $m_i$ and $w_i^*$ for the proposed deep NN. For the output $w_i^*$, an extraction loss that minimizes the difference between $w_i^*$ and $w_i$ is computed to ensure full extraction of the watermark. For the output $m_i$, a fidelity loss that minimizes the difference between $m_i$ and $c_i$ is computed to enable watermark invisibility. For the output $m_i$, we also compute an information loss that forces $m_i$ to contain the information of $w_i$. To achieve this, we maximize the correlation between a feature map of $w_f^i$ and a feature map of $m_i$. Denoting the parameters to be learned as $\vartheta=[\theta_1, \theta_2, \theta_3, \theta_4, \theta_5]$, the loss function $L(\vartheta)$ of the proposed scheme can be expressed as:
\begin{dmath}\label{equ:lossfunc}
    L(\vartheta) = \lambda_1 \| w_i^*-w_i \|_1 + \lambda_2 \|m_i-c_i \|_1  + \lambda_3 \psi(m_i,w_f^i), 
\end{dmath}
% \begin{dmath}
%     L(\vartheta) = \frac{1}{B} \Sigma_{i=1}^B [\lambda_1 \| w_i^*-w_i \|_1 + \lambda_2 \|m_i-c_i \|_1  + \lambda_3 \psi(m_i,w_f^i)], 
% \end{dmath}
% $B$ is the number of training examples,
where $\lambda_i,i=1, 2, 3$ is the weight factor and $\psi$ is a function computing the correlation given as:
\begin{dmath}\label{equ:correlation}
\psi(m_i,w_f^i ) = \frac{1}{2} (\|g(f_1(w_f^i)), g(f_1(m_i)) \|_1 + \|g(f_2 (w_f^i)), g(f_2 (m_i)) \|_1),         
\end{dmath}
where $g$ denotes the Gram matrix that contains all possible inner products. 
% \TODO{I still do not fully get the following sentence. It seems that eq. (2) has only one gram matrix.} 
% \TODO{By minimizing the distance between the two Gram matrices of $f_1$ and $f_2$},
%\TODO{check _1_}
By minimizing the distance between the Gram matrices of the feature maps produced by $f_1$ and $f_2$, we maximize their correlation. To extract the feature maps of $m_i$ and $w_f^i$, the intermediate results ($f_1$ and $f_2$ of the "*" convolution block as shown in 
Figures~\ref{fig:parameters} and~\ref{fig:conv_block}) of two layers are applied (further explained in Section~\ref{subsec:designcomps}).

% \TODO{Who are "these loss terms"? Xin, we need to define the loss here} 
In Eq.~\ref{equ:lossfunc}, each two of the fidelity loss, information loss, and extraction loss terms can be a trade-off for image watermarking--for example, minimizing the fidelity loss term to zero means that $m_i$ is identical to $c_i$. However, in this case, there is no embedded information in $m_i$, thus the extraction of $w_i$ will fail. To allow some imperfectness of the loss terms, the mean absolute error (\ie the L1 norm) is selected to highlight the overall performance rather than a few outliers.

With regularization, the proposed scheme objective is represented as $L(\vartheta)+\lambda_4 P$, where $P$ is the penalty term to achieve robustness as in Eq.~\ref{equ:reg_term}, and $\lambda_4$ is the weight controlling the strength of the regularization term. The deep NN needs to learn the parameter $\vartheta^*$ that minimizes $L(\vartheta)+\lambda_4 P$:

%\TODO{In the following equation, do we need parenthesis since we minimize the entire $L(\vartheta)+\lambda_4 P$ factor?}

\begin{equation}
%\vspace{-0.2cm}
   \vartheta^*=\argmin_{\vartheta} [L(\vartheta)+ \lambda_4 P]. 
%\vspace{-0.2cm}
\end{equation}

In the backpropagation during training, the term $\lambda_1 \| w_i^*-w_i \|_1$ is applied by the components of the architecture in their weight updates, while only $\mu_{\theta_1}$ and $\sigma_{\theta_2}$ apply terms $\lambda_2 \| m_i-c_i \|_1$ and $\lambda_3 \psi(m_i,w_f^i )$ to their weight updates. This enables $\mu_{\theta_1}$ and $\sigma_{\theta_2}$ to encode and embed the information in a way that $\varphi_{\theta_3}$ and $\gamma_{\theta_4}$ are able to extract and decode the watermark.

\subsection{Design of Component NNs}
\label{subsec:designcomps}

\begin{figure*}
    \centering
    \includegraphics[width=0.82\linewidth]{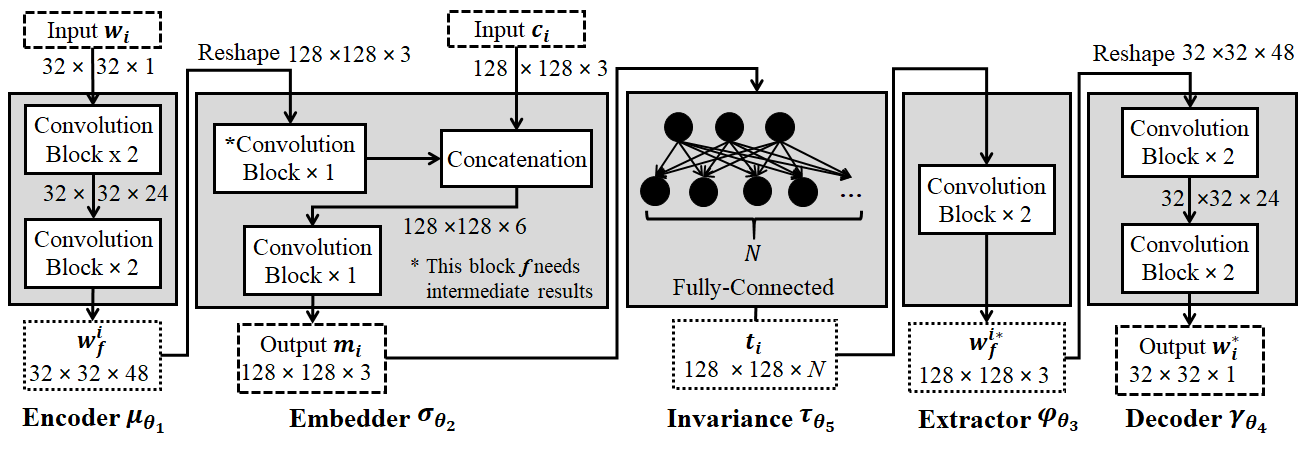}
    %\vspace{-0.4cm}
    \caption{
    Detailed illustration of the components in the proposed watermarking scheme; Encoder $\mu_{\theta_1}$, Embedder $\sigma_{\theta_2}$, the invariance layer $\tau_{\theta_5}$, Extractor $\gamma_{\theta_4}$, and Decoder $\varphi_{\theta_3}$.
    }
    % \TODO{Are all the convolution blocks the same as the * convolution block? ANS: Yes. All of them are the same as the one in Fig.4.}
    %\vspace{-0.5cm}
    \label{fig:parameters}
\end{figure*}

This subsection describes the design of the component NNs $\mu_{\theta_1}$, $\sigma_{\theta_2}$, $\varphi_{\theta_3}$, $\gamma_{\theta_4}$ and $\tau_{\theta_5}$ in more detail. The overall design is modularized and is illustrated in Figure~\ref{fig:parameters}. If we single out two pairs ($\mu_{\theta_1}$, $\gamma_{\theta_4}$) and ($\sigma_{\theta_2}$, $\varphi_{\theta_3}$), we can find that each pair is conceptually symmetrical. 

\subsubsection{The Encoder $\mu_{\theta_1}$ and the Decoder $\gamma_{\theta_4}$ NNs}
Taking the samples $w_i, i=1,2,3,...$ from the input space $W$, the encoder NN $\mu_{\theta_1}$ learns an encoding function that converts $W$ to its feature space $W_f$. Inversely, the decoder NN $\gamma_{\theta_4}$ learns a decoding function from $W_f$ to $W$ with samples $w_f^{i*}, i=1,2,3,...$. $\mu_{\theta_1}$ increases a $32 \times 32 \times 1$ watermark image to $32 \times 32 \times 24$ and $32 \times 32 \times 48$, and $\gamma_{\theta_4}$ successively decreases the $32 \times 32 \times 48$ feature space to a $32 \times 32 \times 1$ watermark image. The reason to train this channel-wise increment is two-fold. First, it produces a $128 \times 128 \times 3$ $w_f^i$ that has the same width and height as the cover image, so that we can concatenate a feature map of $w_f^i$ and $c_i$
%\TODO{check _2_}
% \TODO{Who are "them"? The image samples?} 
along their channel dimension. Each of $w_f^i$ and $c_i$ will contribute equally to the $128 \times 128 \times 6$ concatenated matrix used in the embedder NN $\sigma_{\theta_2}$, thus, we are evenly weighing the watermark and the cover image. Second, the increment in the latent space $W_f$ introduces redundancy, decomposition, and perceivable randomness to $W$, which helps with robustness.

%but also provides additional security.

\subsubsection{The Embedder $\sigma_{\theta_2}$ and the Extractor $\varphi_{\theta_3}$ NNs}
The embedder NN $\sigma_{\theta_2}$ applies the convolution block $f$
% (\TODO{Is this the convolution block of Figure 4? ANS: Correct.}) 
to extract a $128 \times 128 \times 3$ to-be-embedded feature map of $w_f^i$ that is concatenated along the channel dimension with the cover image. Directly applying $c_i$, while only applying a feature map of $w_f^i$, helps $c_i$ to dominate the appearance. The $128 \times 128 \times 6$ concatenation is fed into another convolution block to produce $m_i$. The extractor NN $\varphi_{\theta_3}$ inverses the process by two successive convolution blocks.

\begin{figure}
    \centering
    \includegraphics[width=1\linewidth]{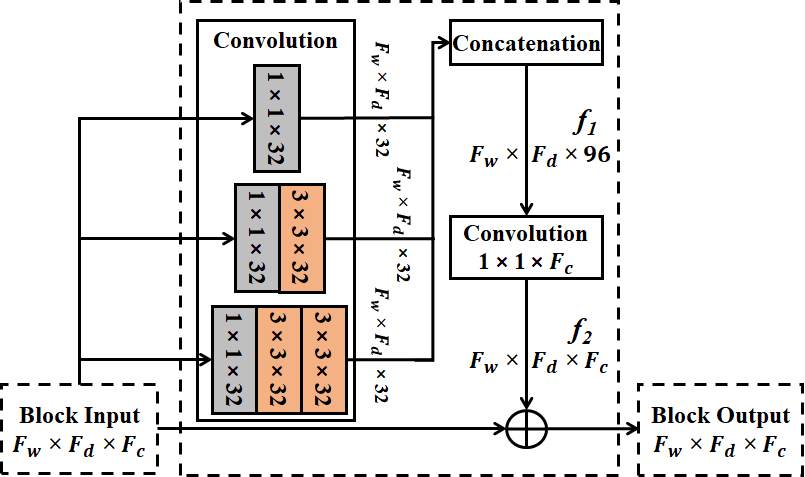}
    \vspace{-0.1cm}
    \caption{Design of a convolution block.}
    %\caption{Convolution block $f$.}
    \vspace{-0.3cm}
    \label{fig:conv_block}
\end{figure}

To capture various scales of features for image watermarking, the inception residual block~\cite{szegedy2017inception} is applied. All the convolution blocks in Figure~\ref{fig:parameters} have the structure shown in Figure~\ref{fig:conv_block}, where $F_w$, $F_d$, and $F_c$ respectively denote the height, width, and the channel of the block input. In the case of the "*" convolution block $f$ of Figure~\ref{fig:parameters}, the annotated intermediate results $f_1$ and $f_2$ of Figure~\ref{fig:conv_block} are applied in Eq.~\ref{equ:correlation}. Specifically, block $f$ extracts features not only from $w_f^i$, but also from $m_i$. The annotated $F_w \times F_d \times 96$ and $F_w \times F_d \times F_c$ feature maps are the intermediate results $f_1$ and $f_2$ respectively.

\subsubsection{The Invariance Layer $\tau_{\theta_5}$}
This is the key component to provide robustness in the proposed image watermarking scheme. Using a fully-connected layer, $\tau_{\theta_5}$ learns a transformation from space $M$ to an over-complete space $T$, where the neurons are activated sparsely. The idea is to redundantly project the most important information from $M$ into $T$ and to deactivate the neural connections of the areas on $M$ irrelevant to the watermark, thus preserve the watermark even if there is noise or distortion that modified a part of $M$. As shown in Figure~\ref{fig:parameters}, $\tau_{\theta_5}$ converts a $3$-channel instance $m_i$ of $M$ into an $N$-channel ($N \geq 3$) instance $t_i$ of $T$, where $N$ is the redundant parameter. Increasing $N$ results in increased redundancy and decomposition in $T$, which provides higher tolerance of the errors in $M$ and thus enhances robustness. 

%~\cite{rifai2011contractive}

Referring to the contractive autoencoder, $\tau_{\theta_5}$ employs a regularization term that is obtained by the Frobenius norm of the Jacobian matrix of its outputs with regards to its inputs. Mathematically, the regularization term $P$ is given as:

%\vspace{-0.3cm}
\begin{equation}
P = \Sigma_{i,j} \left(\frac{\partial h_j(X)}{\partial X_i} \right)^2, 
%\vspace{-0.3cm}
\end{equation}

where $X_i$ denotes the $i$-th input and $h_j$ the output of the $j$-th hidden unit of the fully connected layer. Similar to a gradient computation, the Jacobian matrix can be written as:

\begin{equation}
%\vspace{-0.2cm}
\frac{\partial h_j(X)}{\partial X_i} = \frac{\partial A(\omega_{ji} X_i)}{\partial \omega_{ji} X_i} \omega_{ji},
\end{equation}

where $A$ is an activation function and $\omega_{ji}$ is the weight between $h_j$ and $X_i$. We set $A$ as the hyperbolic tangent ($\tanh$) for strong gradients and bias avoidance~\cite{lecun2012efficient}. Hence, $P$ can be computed as:

\begin{equation}\label{equ:reg_term}
%\vspace{-0.cm}
P = \Sigma_j (1-h_j^2)^2 \Sigma_i (\omega_{ji}^T)^2.
\end{equation}
If the value of $P$ is minimized to zero, all weights $\omega$ in $\tau_{\theta_5}$ will be zero, so that the output of $\tau_{\theta_5}$ will be always zero no matter how we change the inputs $X$. Thus, minimizing P alone will cause the rejection to all the information from the inputs $m_i$. Therefore, we place P as a regularization term in the total loss function to teach $\tau_{\theta_5}$ to preserve useful information related to the loss terms of image watermarking, while rejecting all other noise and irrelevant information. In this way, we achieve robustness without prior knowledge of possible attacks.

%% file: evaluation.tex
In this section, we evaluate our \sol framework by presenting experimental results on: (i) the training and testing of the deep NN; (ii) the overhead that \sol introduces to the IoT onboarding process; and (iii) the robustness of \sol's watermarking scheme.

%We perform detailed experiments to evaluate the watermarking approach for IoT boostrapping, focusing on its ability to secure transmission at different scenarios in real time and evaluate its robustness in real world. We also seek to explore how the setting of cameras affects its performance.

\begin{table*}[htb]
\vspace*{2 mm}
\centering
\begin{tabular}{|l|l|l|l|l|l|l|}
\hline
\multicolumn{1}{|c|}{\textbf{End-to-end onboarding delay (sec)}} & \multicolumn{2}{c|}{\textbf{Deep NN processing time (sec)}}             & \multicolumn{2}{c|}{\textbf{GPU utilization (\%)}}               & \multicolumn{2}{c|}{\textbf{Memory usage (GBs)}}           \\ \hline
                                                    & \multicolumn{1}{c|}{Embedding} & \multicolumn{1}{c|}{Extraction} & \multicolumn{1}{c|}{Embedding} & \multicolumn{1}{c|}{Extraction} & \multicolumn{1}{c|}{Embedding} & \multicolumn{1}{c|}{Extraction} \\ \hline
 \multicolumn{1}{|c|}{$2.53 \pm 0.32$}                                           & \multicolumn{1}{|c|}{$1.12 \pm 0.11$}                         & \multicolumn{1}{|c|}{$1.62 \pm 0.21$}                          & \multicolumn{1}{|c|}{0-26}                      & \multicolumn{1}{|c|}{0-41}                               & \multicolumn{1}{|c|}{7.47}                           & \multicolumn{1}{|c|}{11.21}                               \\ \hline
\end{tabular}
\caption{Experimental IoT onboarding results (average and standard deviation)}
\label{table:results}
\vspace{-0.5cm}
\end{table*}

\subsection{Experimental Setup}

\noindent \textbf{Deep NN deployment, training, and testing:} The deep NN is trained and tested on four NVIDIA TITAN Xp (12GB) GPUs. The watermarking scheme is trained as a single deep NN using the ImageNet dataset~\cite{russakovsky2015imagenet} for cover images and the binary version of the CIFAR dataset~\cite{krizhevsky2009learning} for watermarks, to introduce a large scope of instances to the proposed scheme. 10,000 images from each dataset are separated as the validation set. The testing is performed on 10,000 image samples from the Microsoft COCO dataset~\cite{lin2014microsoft} as the cover image, and 10,000 images of the testing division of the binary CIFAR as the watermark. In Section~\ref{subsec:trainingresults}, we present results on: (i) training and testing of the deep NN; and (ii) the {\em Peak Signal-to-Noise Ratio (PSNR)} and {\em Bit-Error-Rate (BER)}, which are respectively used to quantitatively evaluate the fidelity of the marked image and the quality of the watermark extraction in the testing process. The PSNR is defined as:
\begin{equation}
%\vspace{-0.2cm}
   PSNR =10\log_{10} \left(\frac{(max(c_i)^2}{MSE(c_i, m_i)} \right)
\end{equation}
where $MSE$ is the mean squared error. The BER is computed as the percentage of error bits on the binarization of the watermark extraction $w_i^*$. Finally, in Section~\ref{subsec:rob}, we present results on the robustness in terms of BER for varying marked image distortion percentages in comparison with QR codes.

\noindent \textbf{Mobile application and IoT onboarding:} We used a QR code as the carrier image having a stream of bits (\eg a user key, a password, a secret image, a user's fingerprint scan) as the watermark. The marked image is another QR code. We have developed a prototype Android application, so that users can take pictures of marked images (QR codes). These pictures are sent over WiFi to the GPU, where the watermarks are extracted and the user credentials are verified. If the verification is successful, the IoT device receives a url that will take it to a server for onboarding. In cases that the user's fingerprint scan has been embedded as the watermark, the user device needs to be equipped with a fingerprint scanner. The user will be asked to scan their fingerprint, since this fingerprint scan will be sent to the server, which will verify that it matches the scan embedded into the QR code. In Section~\ref{subsec:networkingresults}, we present results on the following metrics:

\begin {itemize}[leftmargin=0cm,itemindent=.3cm,labelwidth=\itemindent,labelsep=0cm,align=left, noitemsep, topsep=0pt]

\item \emph {End-to-end delay for onboarding:} This is measured as the time elapsed between the moment that a user takes a picture of a marked image and sends it to the watermark extraction deep NN until the moment that the IoT device is onboarded. 

\item \emph {Processing time for watermark embedding and extraction:} The processing time needed by the deep NNs to embed a watermark into a cover image and to extract a watermark from a marked image after it is requested by a user.

\item \emph {Utilization of resources during deep NN operation:} The GPU and memory usage during watermark embedding and extraction.

\item \emph {Accuracy of biometrics-based \sol:} The \sol accuracy when the users' fingerprint scans are used for onboarding given that the scans embedded into the QR code and the scans sent to the server may be similar, but not identical.

\end {itemize}

\subsection {Experimental Results}
%\vspace{-0.2cm}

\subsubsection{NN training and testing}
\label{subsec:trainingresults}

\begin{figure}[t]
%\vspace{-0.2cm}
\centering
\begin{subfigure}{1\linewidth}
 \centering
 \includegraphics[width=0.65\textwidth]{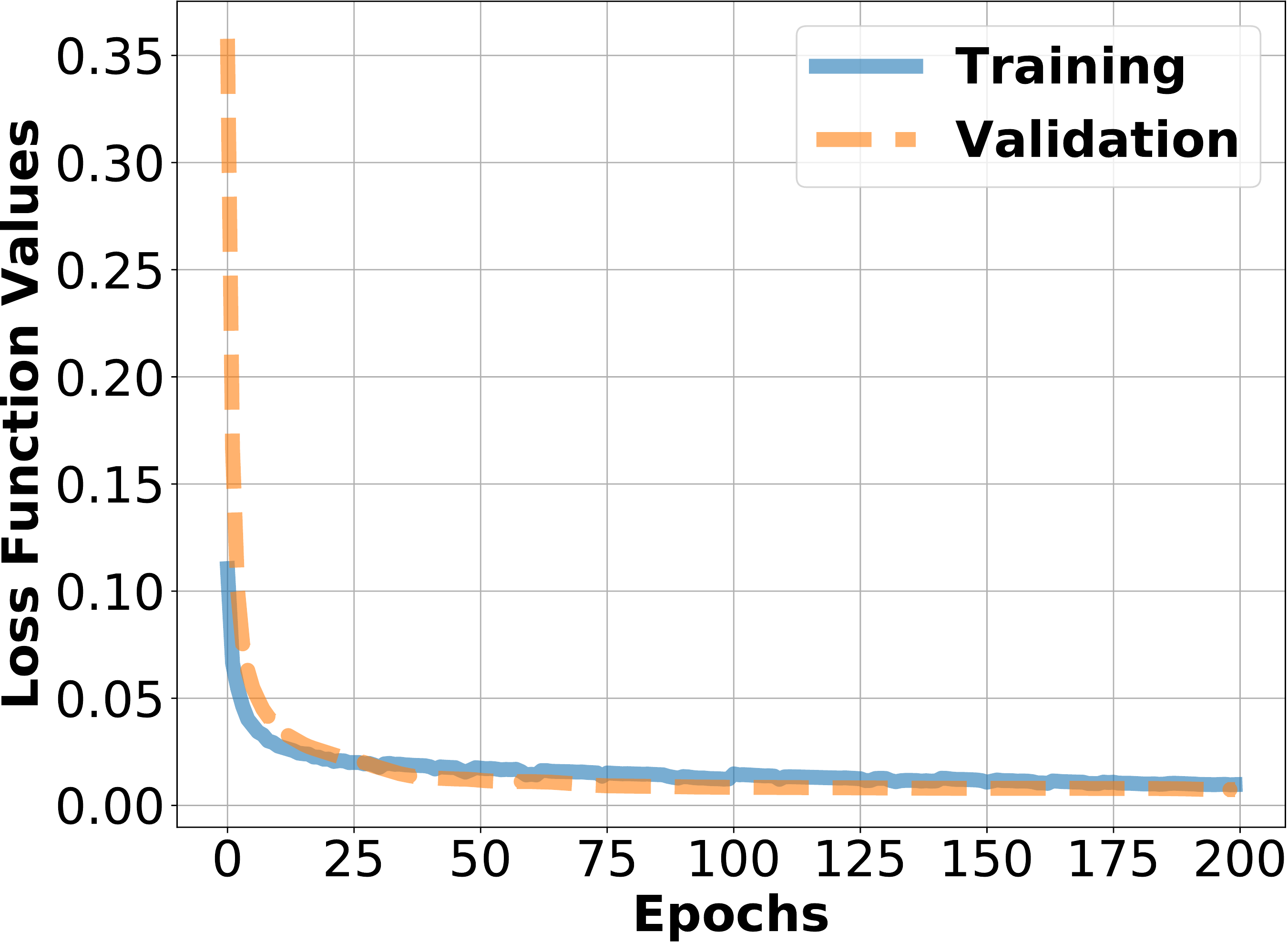}
% \vspace{-0.45cm}
\caption{Training results: the NN loss function during 200 epochs.}
  \label{fig:results}
 \end{subfigure}
 \begin{subfigure}{1\linewidth}
 \vspace{0.2cm}
 \includegraphics[width=0.65\textwidth]{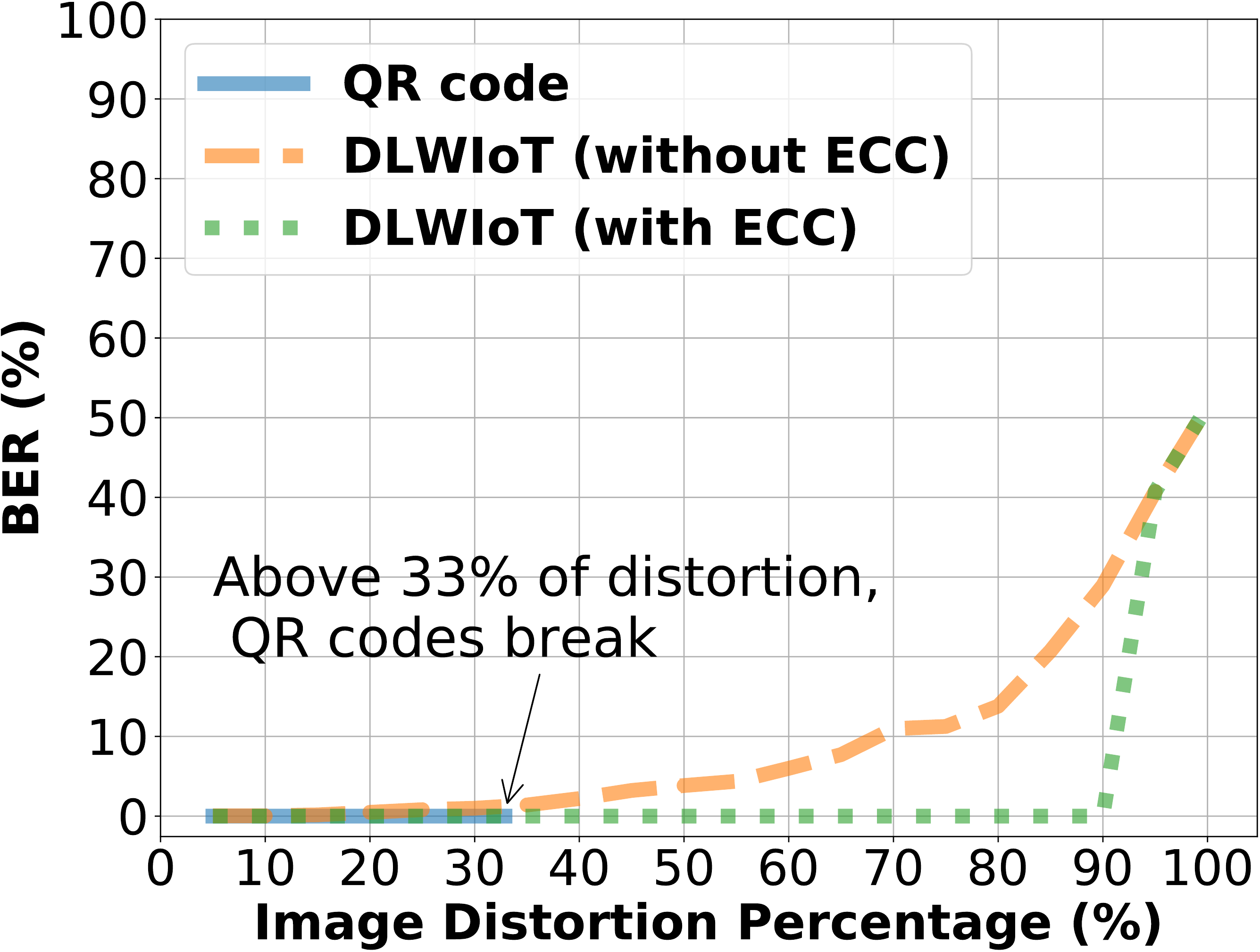}
  %\vspace{-0.45cm}
  \centering
  \caption{BER for varying image distortion percentages.}
  \label{fig:robustness}
 \end{subfigure}
  \vspace{-0.1cm}\caption{Deep NN training and watermark robustness results.}
 \vspace{-0.5cm}
\end{figure}

Figure~\ref{fig:results} illustrates the NN loss function $L(\vartheta)+\lambda_4 P$ during 200 epochs. During the training and validation, the value of $L(\vartheta)+\lambda_4 P$ converges below 0.03, indicating a proper fitting.
% where we define $T1=\lambda_2 \|m_i-c_i \|_1  + \lambda_3 \psi(m_i,w_f^i)$ and $T2=\lambda_1 \| w_i^*-w_i \|_1$. At the training and validation, both $T1$ and $T2$ in converge smoothly below 1.5\% and $L(\vartheta)+\lambda_4 P$ converges below 3\%, indicating a proper fitting of the training.

In testing, the BER is zero, indicating that the original and the extracted watermarks are identical. The testing PSNR is $39.72 dB$, indicating a high fidelity of marked images, so that the hidden information cannot be identified by human vision.

%\begin{figure}
%    \centering
%    \includegraphics[width=0.6\linewidth]{figs/loss-function.pdf}
%    \caption{Training results: The NN loss function during 200 epochs.}
%    \label{fig:results}
%    \vspace{-0.5cm}
%\end{figure}

\subsubsection{IoT onboarding}
\label{subsec:networkingresults}

We present our results in Table~\ref{table:results}.

\noindent \textbf{End-to-end onboarding delay:} This delay consists of the network delay to/from the GPU that runs the watermark extraction deep NN ($54msec$ roundtrip delay), the extraction deep NN processing time, the processing by the user mobile phone, the network delay to the onboarding server ($52msec$ roundtrip delay), and the processing by the onboarding server. Our results indicate that the end-to-end onboarding delay is roughly $2.5sec$. Even in cases of longer roundtrip delays (\eg distant clouds with roundtrip delays of $150-200msec$), the onboarding delay is not expected to exceed $3sec$.

\noindent \textbf{Processing delay for watermark embedding and extraction:} Our results indicate that embedding by the deep NN lasts about $1.12sec$. The extraction process lasts $1.62sec$, requiring about $1.5x$ more processing time than embedding.

\noindent \textbf{Resource utilization during deep NN operation:} The watermark embedding and extraction processes utilize a single GPU up to $26\%$ and $41\%$ respectively. The memory consumption is $7.47GB$ for embedding and $11.21GB$ for extraction.

\noindent \textbf {Accuracy of biometrics-based \sol:} The accuracy of \sol reaches $99\%$. The onboarding and processing delays and the resource utilization results reported above still hold.

%Note that currently our NNs and experimental setup allow for up to 2 operations (watermark embedding and/or extraction) being executed simultaneously per GPU. Incoming user operation requests are buffered until computing resources become available. 

\subsubsection{\sol Watermark Robustness}
\label{subsec:rob}

In Figure~\ref{fig:robustness}, we present the BER for varying marked image distortion percentages for \sol and legacy QR codes. The results demonstrate that legacy QR codes can tolerate distortions up to 1/3 of the image through the application of Error Correction Coding (ECC). However, \sol can tolerate distortions of up to 2/3 of the marked image without ECC, while with ECC, \sol can tolerate distortions up to 85-90\%.

%\begin{figure}
%    \centering
%    \includegraphics[width=0.6\linewidth]{figs/robustness.png}
%    \caption{BER for varying image loss percentages.}
%    \label{fig:robustness}
%    \vspace{-0.5cm}
%\end{figure}

%\subsection{}

%\subsection{Performance Comparison}
%\TODO{Here, we need some overhead estimation for our watermarking approach for IoT onboarding compared to using a baseline approach for IoT onboarding with a plain QR code. We also need some accuracy numbers for the model.}

%\subsection{Processing Time}
%\TODO{Peggy, I recall you had another idea on what to evaluate?}
%The application is implemented in our test hardware.

%% file: related.tex
%\TODO{Xin, can you please add some related work on ML/deep-learning for image watermarking here? I will also add some related work on IoT bootstrapping}

\noindent \textbf{Related Work on IoT Onboarding:} The selection of onboarding techniques depends on the design of the security architecture (\eg distributed, centralized). Out-of-band techniques include the use of QR code and/or pre-defined passwords by users~\cite{latvala2020evaluation}. Onboarding in centralized architectures often relies on pre-established trust relations and utilize protocols, such as Extensible Authentication Protocol (EAP)~\cite{vollbrecht2004extensible}, for authentication. In distributed architectures, devices do not rely on pre-established trust relations--onboarding results in credentials being created for security in subsequent communication. To this end, peer IoT devices can perform a Diffie-Hellman type of handshake to agree on a common secret~\cite{diffie1976new}. Protocols such as IKEv2~\cite{kaufman2005internet} and TLS~\cite{dierks2008transport} allow peers to exchange keys and establish security associations without the need for a connection to a trusted server. Furthermore, approaches to secure and automate IoT onboarding through trusted hardware have been proposed~\cite{intelintel}. In this paper, we focused on a centralized architecture through a deep learning-based watermarking scheme for IoT onboarding.

\noindent \textbf{Related Work on Deep Learning for IoT:} The potential of deep learning in the context of IoT has been discussed in prior work~\cite{li2018learning, al2018survey}. Deep learning has been used to detect tampered IoT devices~\cite{ferdowsi2018deep}, while work has also been done on running deep learning on IoT devices~\cite{tang2017enabling}. In this paper, we focused on deep learning to enable the onboarding of IoT devices by authorized users through a watermarking scheme.

% verhelst2017embedded

\noindent \textbf{Related Work on Deep Learning for Image Watermarking:} Although still at its infancy, incorporating deep learning into image watermarking has attracted increased attention in recent years. Zhong \textit{et al.}~\cite{zhong2020automated} investigated a general-purpose deep learning-based watermarking scheme without considering the embedding of user biometrics to cover images. Kandi \textit{et al.}\cite{kandi2017exploring} used two deep autoencoders to indicate bits 1 and 0 respectively for a non-blind binary watermark extraction. By embedding via adversarial images and extracting through the first layer of a deep NN, Vedran \textit{et al.}\cite{vukotic2018deep} developed a single-bit watermarking scheme. %Li \textit{et al.}\cite{li2019novel} embedded a watermark in the discrete cosine domain of a cover image and applied convolutional NNs to assist the extraction. 
In scenarios where a master share is sent separately from the image, Fierro-Radilla \textit{et al.}\cite{fierro2019robust} linked the watermark with features from the cover image extracted by convolutional NNs to create the master share. Due to the fragility of deep NNs \cite{papernot2016limitations}, robustness is a challenge, since noise or modification on the marked image can destroy the trained models. Mun \textit{et al.}\cite{mun2019finding} proposed to solve this issue by enumerating adversarial examples during training. In this paper, we achieved robustness without adversarial examples of potential attacks and tolerate noise on the marked images without requiring any information from the cover images.

%\xin{Different from the previous approaches, the presented image watermarking scheme is the first method that applies deep NNs to reduce the requirement of domain knowledge, while tolerating noise on the marked images without requiring any information from the cover images.}
%QR code based watermarking~\cite{thomas2018multilevel}

%Mun \textit{et al.}\cite{mun2019finding} proposed to solve this issue  by proactively including  adversarial  attacking examples during  the  training. But it is impossible to enumerate all the attacks in practice, and hence this method has a limited range of applications.

%% file: conclusion.tex
In this paper, we presented \sol, a framework for IoT onboarding through image watermarking. In \sol, user credentials can be covertly embedded into an image (\eg a QR code), which can be printed on an IoT device and can be used only by an authorized user %(\ie the user with the embedded credentials) 
to onboard the device. In its core, \sol features a novel deep learning-based scheme for image watermarking that offers robustness against distortion and quality degradation of marked images. %(\eg when the user takes a picture of the device QR code with his/her mobile phone).

While \sol is off to a promising start, the current framework supports the embedding and extraction of the credentials of a single user. As a result, an IoT device can have a single owner, who can onboard it. In our future work, we plan to enable the embedding and extraction of the credentials of a group of users, so that IoT devices can be onboarded by a group of authorized users. We also plan to tackle changes of device ownership--for instance, cases of device re-selling, where the device owner sells the device to another individual. 